\documentstyle[aaspp4]{article}
\begin{document}
\title{Heavy neutrino dark matter in the solar system}
\author{Faustin Munyaneza  
 and Raoul D. Viollier\footnote{E-mail:  viollier@physci.uct.ac.za}}
\affil{Institute of Theoretical Physics and Astrophysics, \\
 Department of Physics, University of Cape Town, \\
 Rondebosch 7701, South Africa}

\begin{abstract}
We study  a
simple model of dark matter  that is gravitationally clustered around the
sun in the
 form of
a spherical halo of
a degenerate gas of heavy neutrinos.
It is shown that for neutrino masses  $m_{\nu} \stackrel {\textstyle <}{\sim}
 16~{\rm keV}/c^{2}$, the resulting  matter distribution is consistent with
the constraints on 
the  mass excesses
within the orbits of the outer planets, as obtained from  astrometrical
and the Pioneer 10/11 and Voyager 1/2 (Anderson et al. 1995)
ranging data.
However, the  anomalous acceleration
recently detected in the Pioneer 10/11 data that is approximately constant
between 40 AU and 60 AU (Anderson et al. 1998; Turyshev et al. 1999) is
incompatible with both our model
and earlier Pioneer 10/11 ranging data for the outer planets.
We then calculate the planetary and asteroidal perihelion shifts generated
by
 such a neutrino halo.
For $m_{\nu} \stackrel{\textstyle <}{\sim}16~{\rm keV}/c^{2}$, our
 results are consistent
with the observational data on Mercury, Venus, Earth and Icarus.
Finally, we propose to detect  this neutrino halo
 directly  with a dedicated
experiment on Earth,
observing the X-rays emitted in the radiative decay of the heavy neutrino into
 a light neutrino and a photon.

\end{abstract}
\keywords{Gravitation-Dark matter-Solar system:general}

\section{Introduction}
It is well known that the properties of galactic systems pose
a great challenge to gravity theories.
For virtually all spiral galaxies the
galactic rotation curves
 tend towards
some constant value for large distances from the center of the galaxy.
This is clearly either inconsistent with the mass distribution
inferred from
the distribution of visible stars, or, even more challenging,
 in contradiction with the laws of Newtonian dynamics.

Most widely accepted is the conservative  explanation of this discrepancy
in terms of dark matter (DM) 
(e.g. Ashman 1992) which 
 presumes that the visible stars are embedded in a massive, nearly
spherical halo
of nonluminous matter.
The mass of the halo
varies from one galaxy to another,
but in general  it constitutes 90\% of the total mass (Tremaine 1992).
While the DM hypothesis could explain the flat rotation curves of the 
galaxies in a consistent manner,
 it has its own troubles, in particular:
(i) there is  no compelling model for the formation of the DM halo,
 and (ii) despite much effort, so far no known form of matter
lends itself
to a satisfactory understanding of the DM halo. For instance, the  microlensing
experiments MACHO (Alcock et al. 1996, 1997) and EROS (Ansari et al. 1996),
are as yet
far from explaining  the galactic DM halo  in terms 
of massive  astrophysical compact halo objects (MACHOs). 
In fact, it is by no means
clear that the DM halo should be made
of astrophysical low-mass black holes, brown dwarfs, old white dwarfs or
more exotic nonbaryonic MACHOs.
On the contrary, the DM halo may very well consist of loose clouds
of weakly interacting (i.e. nonbaryonic) massive particles (WIMPs)
such as SUSY neutralinos, neutrinos or axions (Jungman et al. 1996).
Microlensing experiments have ruled out a large class
of possible compact baryonic DM components.
As gaseous baryonic components are also largely
excluded, it has been argued that  at least some nonbaryonic DM is required
 to explain  the DM halo in our Galaxy (Freese et al. 1999).
Of course, if DM is in the form of
nonbaryonic particle clouds, it will be everywhere in our Galaxy
and it may thus also be observed in the solar system.

An alternative explanation for the galactic rotation curves is based
on the possibility
that Newton's gravitational law ceases to be valid at small accelerations
(Milgrom 1983; Bekenstein 1992) or large scales (Sanders 1990).
This challenging hypothesis
has been discussed in a number of papers 
(e.g. Eckhardt 1993; Hammond 1994).
For instance, it has been shown 
 (Sanders  1986; Nieto and Goldman 1992)
that  a modified gravitational potential of the form
\[ \phi=\frac{-GM}{r(1+\alpha)}\left[1+\alpha {\rm e}^{-r/r_{0}}\right] , \]
with $\alpha=-0.9$ and  $r_{0}\approx 30$ kpc, could indeed
 explain the flat rotational
curves for most
of the galaxies.
In contrast to the DM hypothesis,
 there would be no detectable corrections to
Newton's gravity,
as the exponential term is unity with high 
accuracy for the solar system.
We would  thus recover the standard form for
the gravitational potential $\phi=-GM/r$ within the solar system.
The problem with this hypothesis is that
it is not clear how to incorporate the modified gravitational
potential in a fully general relativistic theory.

The main motivation  for studying   DM in the solar system 
is that galaxies could have halos consisting of 
 weakly interacting massive particles (WIMPs) ( Trimble 1987). 
Indeed, if DM exists in the
form of WIMPs, one would expect  this exotic  form of DM to
penetrate both the galactic disk and 
the solar system.
The determination of the total matter
density in the vicinity
of the sun is a classical problem of  astrophysics (see Tremaine (1990)
and Oort (1965) for a comprehensive review of early research).
It allows for 
 a significant
amount of DM that  could have  condensed
into a DM halos
around the stars.
However, the  conditions
on particle masses and
interaction
cross sections under which this
condensation would have taken place during star formation,
 have not been
worked out in detail. In fact, it is  difficult to
envisage  a gravitational interaction mechanism
 that would lead to capture of
significant amounts of weakly interacting DM particles.                                                                                                  
Assuming these particles to be 
dissipationless, they would be part
of the galactic halo and move with the
velocities of the order of $10^{-3}c$. In this case
 very few DM particles  could be trapped
in orbits around the sun
and the DM density would be nearly constant near the sun.
 However, Bahcall (1984a, 1984b) has argued for the
 possible existence
of a galactic  disk component of nonluminous matter,
not necessarily weakly interacting, with a density of the order of
luminous disk matter, i.e.  $0.1 \ M_{\odot} {\rm pc}^{-3}$.
Such DM could have been trapped around  the sun since its formation.
In fact,
it is possible that ``nonequilibrium'' gravitational
interaction between a cloud of   weakly interacting DM particles and 
a molecular cloud, could 
play an important role in the formation of stars with DM halos
around them. The obvious place to test such a hypothesis is
the solar system.
In fact,
Mikkelsen and Newman (1977) established a bound of
 $M_{d} \stackrel {\textstyle <}{\sim} 7 \times10^{-8}M_{\odot}$ for
DM within  the Earth's orbit, while
 Khloper et al. (1991)  made  estimates
of $10^{-7}$  to $10^{-6} M_{\odot}$ of shadow matter
being captured by a baryonic matter star.
If a DM halo  exists around the sun, it would induce an additional
perihelion precession  of the  planetary orbits (Braginsky et al. 1992).
More recently,
Anderson et al. (1995) have investigated the bounds on DM 
within the orbits of different outer planets using the data from
Pioneer 10/11 and Voyager 1/2. They found, 
within Uranus' orbit, an upper bound for the DM mass of
$M_{d} \stackrel{\textstyle <}{\sim} 0.5 \times 10^{-6}M_{\odot}$, while 
 a bound 
of  $M_{d} \stackrel{\textstyle <}{\sim}3 \times 10^{-6}M_{\odot}$ was
 obtained within Neptune's orbit, and finally, 
within Jupiter's orbit, the DM mass was determined  to
 $M_{d}= (0.12\pm 0.027)\times 10^{-6}M_{\odot}$.

Thus, in this paper, we  assume that massive neutrinos play the role
of the weakly interacting DM particles, that have been trapped around
the sun during its formation, and we  investigate the
consequences of the existence of  such a degenerate neutrino halo.
In fact, the  idea  that DM is made of massive
 neutrinos that can cluster around a baryonic star has
been developed in a series of papers (Viollier  et al. 1992, Viollier
1993; Viollier 1994). This model resembles
the Thomas-Fermi model   of an atom or  ion,
where the degenerate electron cloud
around the nucleus is described by a mean   electrostatic field.
The difference between the two models is the sign
of the interaction.
While the model of the neutrino halo around  a star admits also 
solutions without star at the center, 
 i.e. pure neutrino balls (Viollier
1994),
 there is no corresponding  solution
for a nucleus with charge $Z=0$. Here it is important to
note that  
pure neutrino balls could  also  describe  the  supermassive 
compact dark objects 
 at the galactic
centers, e.g. at the center of our Galaxy or M87
 (Tsiklauri and Viollier 1998;
 Bili\'c, Munyaneza and Viollier 1999;
 Munyaneza, Tsiklauri and Viollier 1998, 1999; Munyaneza and Viollier 1999).
   
The purpose of this paper is to study
the mass distribution of degenerate neutrino
 DM around the sun and to  compare it
with the most recent observational data from Pioneer 10/11 and Voyager 1/2
(Anderson et al. 1995). We will also  investigate
the effect of a  possible  neutrino halo  on the precession of the
perihelion of  planetary or asteroidal orbits.

This paper is organized as follows:
In section~2, we  establish the basic equations of our  model and study
the mass distribution of the neutrino halo around the sun.
 In section~3, we investigate the
planetary and asteroidal  perihelion shifts due to a  neutrino halo.
We discuss other  observational implications of  such a neutrino halo
in section~4 and 
 summarize our results with a discussion in section~5.

\section{A degenerate heavy neutrino halo around the sun}

Let us  characterize the  spherically symmetric cloud of
degenerate heavy neutrino matter around  a baryonic star by
its
gravitational potential $\Phi(r)$,
 pressure $P_{\nu}(r)$, 
and  mass density $\rho_{\nu}(r)$.
In nonrelativistic approximation,
these three quantities are linked through 
Poisson's equation
\begin{equation}
\Delta \Phi =4\pi G \rho_{\nu} ,
\label{eq:101}
\end{equation}
the condition for hydrostatic equilibrium
between the gravitational and degeneracy pressures of heavy neutrino matter
\begin{equation}
\frac{dP_{\nu}}{dr}=-\rho_{\nu} \frac{d\Phi(r)}{dr},
\label{eq:102}
\end{equation}
and the polytropic equation of state of a degenerate
nonrelativistic Fermi gas
\begin{equation}
P_{\nu}=K\rho_{\nu}^{5/3} \ .
\label{eq:103}
\end{equation}
Here the constant  $K$ is given by
\begin{equation}
K=\left(\frac{6}{g_{\nu}}\right)^{2/3}
\frac{\pi^{4/3}\hbar^{2}}{5m_{\nu}^{8/3}} \  ,
\label{eq:104}
\end{equation}
 $g_{\nu}$ being  the spin degeneracy factor of the  neutrinos and
 antineutrinos, i.e. $g_{\nu}=2$ for Majorana neutrinos and
$g_{\nu}=4$ for Dirac neutrinos and antineutrinos.
The pressure and density vanish at 
the radius $R_{0}$ of the neutrino halo, i.e.
\begin{equation}
\label{eq:105}
\rho_{\nu}(R_{0})=P_{\nu}(R_{0})=0 \  .
\end{equation}
Outside the neutrino halo, $r > R_{0}$, we thus recover
 the standard solution of  Poisson's equation
(\ref{eq:101})
\begin{equation}
\label{eq:106}
\Phi (r)=-\frac{G(M_{\nu}+M_{B})}{r} \ ,
\end{equation}
where $M_{\nu}$ is the  mass of the neutrino halo, i.e.
\begin{equation}
M_{\nu}=\int_{0}^{R_{0}} 4\pi \rho_{\nu}(r)r^{2}dr ,
\label{eq:107}
\end{equation}
 and $M_{B}$ is the mass of the pointlike baryonic star
at the center of the neutrino halo.

Introducing  the   dimensionless
potential variable $v$ and radius $x$, by 
\begin{equation}
v(x)=\frac{r}{GM_{\odot}}\bigg(\Phi(R_{0})-\Phi(r)\bigg),
\label{eq:108}
\end{equation}
\begin{equation}
\label{eq:109}
x=\frac{r}{a_{\nu}} ,
\end{equation}
where
  $a_{\nu}$  is an appropriate  length scale 
\begin{equation}
\label{eq:110}
a_{\nu}=
\left(\frac{3\pi\hbar^{3}}{4\sqrt{2}m_{\nu}^{4}g_{\nu}G^{3/2}M_{\odot}^{1/2}}\right)^{2/3}
=
0.4129~{\rm pc}\left(\frac{17.2~{\rm keV}}{m_{\nu}c^{2}}\right)^{8/3}
\left(\frac{2}{g_{\nu}}\right)^{2/3},
\end{equation}
 we arrive at the nonlinear Lan\'e-Emden differential
equation (Viollier, Leimgruber and Trautmann 1992, Viollier 1994)
with  index $n=3/2$
\begin{equation}
\frac{d^{2}v}{dx^{2}}=\frac{-v^{3/2}}{x^{1/2}}, 
\label{eq:111}
\end{equation}
subject to  the boundary conditions 
\begin{equation}
\label{eq:112}
v(0)=\frac{M_{B}}{M_{\odot}} \  \ \  {\rm and} \ \ \ v(x_{0})=0,
\end{equation}
where $x_{0}=R_{0}/a_{\nu}$ is the radius of the halo
in units of $a_{\nu}$.
The first condition in (\ref{eq:112})  arises from the fact
that, near the origin, the potential energy is dominated by $M_{B}$.
Thus $M_{B}=0$ corresponds to a pure neutrino ball without
a pointlike source at the center. The properties of such neutrino balls
have been discussed
in a number of papers (Tsiklauri and Viollier 1998;
 Munyaneza,Tsiklauri and Viollier 1998,1999; Munyaneza and Viollier 1999).
In particular, it has been shown that a neutrino ball
of $M \approx 2.6  \times 10^{6}M_{\odot}$ is a viable
alternative to the   supermassive
black hole  that many believe to exist at  the Galactic center, as
(i) it is consistent with the upper limit
of the size of the supermassive compact dark object
as determined by the  motion of stars near Sgr A*
(Munyaneza, Tsiklauri and Viollier 1998, 1999), and
(ii) the bulk part of its infrared to radiowave spectrum, up to
wavelengths of $\lambda= 0.3~{\rm cm}$, can be
described by standard thin accretion 
disk theory (Bili\'c and Viollier 1998; Bili\'c, Tsiklauri and Viollier
1998; Tsiklauri and Viollier 1999; Munyaneza and Viollier 1999).

The  neutrino matter density 
may  be expressed in terms of the potential variable $v$ as
\begin{eqnarray}
\label{eq:112a}
\rho_{\nu}(r)&=&\frac{8m_{\nu}^{8}g_{\nu}^{2}G^{3}M_{\odot}^{2}} 
{9\pi^{3}\hbar^{6}}\left(\frac{v}{x}\right)^{3/2} \nonumber\\
&=&1.12912 M_{\odot} {\rm pc}^{-3}
\left(\frac{m_{\nu}c^{2}}{17.2~{\rm keV}}\right)^{8}
\left(\frac{g_{\nu}}{2}\right)^{2}
\left(\frac{v}{x}\right)^{3/2}.
\end{eqnarray}
Using Eqs.~(\ref{eq:107}) and (\ref{eq:112a}),
the neutrino matter mass   enclosed
within  a radius $r$  from the sun can be written as 
\begin{equation}
{\cal M}_{\nu}(r)=-M_{\odot}\bigg(xv'(x)-v(x)+v(0)\bigg) .
\label{eq:113}
\end{equation}
with 
\begin{equation}
{\cal M}_{\nu}(R_{0})=-M_{\odot}\bigg(x_{0}v'(x_{0})+v(0)\bigg)=M_{\nu}.
\label{eq:113ab}
\end{equation}

Of course, the mass density $\rho_{\nu}$ diverges
at the origin as $x^{-3/2}$ due to the
pointlike nature of the baryonic  source. However,  the integral in   
Eq.~(\ref{eq:107})
converges, yielding a finite  total mass of the neutrino halo
$M_{\nu}$.
In Fig.~1 we plot the mass ${\cal M}_{\nu}(r)$ that  is  enclosed
 within a radius $r$
from the sun, for various values of the neutrino mass $m_{\nu}$.
The slope of $v$ at the center has been fixed to  $v'(0)=1$,
yielding a total mass of the neutrino  halo of  
 $M_{\nu}=0.7~M_{\odot}$.
Of course, by varying $v'(0)$,  the halo could have any
mass.
Also shown in Fig.~1 is  the bound on DM
mass within the Earth's orbit, i.e. 
$M_{d} \stackrel {\textstyle <}{\sim}7\times 10^{-8}M_{\odot}$ 
(Mikkelsen and Newman 1977) as well as
 the constraints
on the  mass excesses
within the orbits of various outer planets,
as obtained from astrometrical 
and  the Voyager 1 and 2 and Pioneer 10 and 11 ranging data
(Anderson et al. 1995). According to these data
 the DM mass contained
within Jupiter's orbit is $M_{d}=(0.12 \pm 0.027)\times 10^{-6}M_{\odot}$,
within  Uranus' 
orbit $M_{d} \stackrel {\textstyle <}{\sim} 0.5\times 10^{-6}M_{\odot}$,
and  within Neptune's orbit 
 $M_{d} \stackrel {\textstyle <}{\sim} 3\times 10^{-6}M_{\odot}$.
A DM mass of a few $10^{-4}M_{\odot}$ over the range from
$40~{\rm AU}$ to $60~{\rm AU}$ as shown in Fig.~1 would be 
 consistent with the anomalous
acceleration of $a_{P} \approx 7.5 \times 10^{-8} {\rm cm/s^{2}}$
observed  in the Pioneer 10/11, Galileo and Ulisses data (Anderson
et al. 1998; Turyshev et al. 1999).

Based on our model, we  can determine the neutrino-mass range
compatible with the limits on DM for the various planetary orbits.
The constraints on the DM mass
within Earth's orbit yields the  neutrino mass limits 
\begin{eqnarray}
&m_{\nu}c^{2}& \leq 21.8~{\rm keV}  \ {\rm for} \ g_{\nu}=2
\nonumber \\
&m_{\nu}c^{2}& \leq  18.3~{\rm keV} \ {\rm for} \ g_{\nu}=4.
\label{eq:114a}
\end{eqnarray}
Taking the DM data  within Jupiter's orbit at face value and interpreting
DM as degenerate neutrino matter,
the neutrino mass limits are
\begin{eqnarray}
12.8 \ {\rm keV} \leq &m_{\nu}c^{2}& \le 14.2 \  {\rm keV} \ {\rm  for} \  g_{\nu}=2,
\nonumber \\
 10.8 \ {\rm keV} \leq &m_{\nu}c^{2}& \le 11.9 \ {\rm keV} \ {\rm for} \  g_{\nu}=4.
\label{eq:114b}
\end{eqnarray}
However, one  should  perhaps interpret this 
range of  neutrino masses
 as a lower limit,
 as Jupiter tends to eject
any matter
within its orbit (Anderson et al. 1995).

For DM within Uranus' orbit, the upper limits for $m_{\nu}$ are
\begin{eqnarray}
&m_{\nu}c^{2}& \le 12 \  {\rm keV} \  {\rm for} \  g_{\nu}=2, \nonumber\\
&m_{\nu}c^{2}& \le  10 \  {\rm keV} \ {\rm  for} \  g_{\nu}=4,
\label{eq:114c}
\end{eqnarray}
and finally for  Neptune's orbit, the bound on the neutrino mass  is 
\begin{eqnarray}
&m_{\nu}c^{2}& \le 16 \  {\rm keV} \  {\rm for} \  g_{\nu}=2, \nonumber\\
&m_{\nu}c^{2}& \le  13.5 \  {\rm keV} \ {\rm  for} \  g_{\nu}=4.
\label{eq:114d}
\end{eqnarray}

We now explore what the total mass of the neutrino halo  
 should be, in order to be consistent  with the observational 
data.
For this purpose, we have calculated
the mass ${\cal M}_{\nu}(r)$ that is  enclosed within a radius $r$, for
various total masses $M_{\nu}$  of the halo.
The results are shown in Fig.~2, where 
 the neutrino mass is fixed to $m_{\nu}=14~{\rm keV}/c^2$.
As suggested by this figure,
the neutrino halo cannot be heavier than $\sim 10^{2} M_{\odot}$,
 otherwise it would be
inconsistent with the constraints on  the 
 observed mass excesses within the orbits
 of the outer planets.
In Fig.~3, the  acceleration $a_{\nu}$ due to 
the presence of a neutrino halo
is plotted as a function of the radius.
Also shown is the range
of allowed values for the acceleration at the positions of various planets
(Anderson et al. 1995).  
We thus conclude that  our model is consistent
with the observations of the outer planets (Anderson et al. 1995).
However, it cannot explain the apparent anomalous
 weak long-range
acceleration seen in the  Pioneer 10/11, Galileo, and 
Ulysses data (Anderson et al. 1998; Turyshev et al. 1999)
indicated with  a horizontal bar in Fig.~3. These groups
claim that there is  an anomalous acceleration 
towards the sun  of $a_{P} \approx 7.5 \times 10^{-8} {\rm cm/s^{2}}$.
No  variation of $a_{P}$ with
distance from the sun  was found over a range of 40 to 60 AU.
Of course, our neutrino halo would
also contribute
to the anomalous acceleration $a_{P}$.
Setting $a_{P}=a_{\nu}$, our model predicts
a decrease of the anomalous acceleration due to the fact
that the neutrino halo
mass enclosed within  a distance $r$ from   the sun scales as $r^{3/2}$.
However, if the anomalous acceleration
is indeed constant, as the recent observations
(Anderson et al. 1998; Turyshev et al. 1999) seem to indicate,
the enclosed mass should increase like $r^{2}$ which is clearly
inconsistent with the astrometrical and previous ranging data.
Of course, there have been some attempts to explain the Pioneer
anomalous acceleration (Katz 1998, Murphy 1998),
but it is perhaps fair to say that the jury  on the
explanation of this real or spurious effect
 is still out.
In Fig.~4, we present the   mass $M_{\nu}$   as a function
of the radius  $R_{0}$ of the neutrino halo.
It is interesting to note
 that there is always  maximal radius of a halo
around a baryonic star.
For $M_{B}=M_{\odot}$, $m_{\nu}=14~{\rm keV}/c^{2}$ and $g_{\nu}=2$,
 this maximal radius turns out to be $R_{max} \approx 5~{\rm lyr}$,
which is  the scale of the interstellar distances.
 The corresponding neutrino halo mass
would be $M_{\nu ,max} \approx 3 M_{\odot}$.

\section{Planetary and asteroidal perihelion shifts }

A possible neutrino halo will, of course, affect
the perihelion shifts of the planets and asteroids of the solar system.
Using Eq.~(\ref{eq:108}), the gravitational potential $\Phi(r)$ of a sun
that is immersed in  a neutrino halo can be written as
\begin{equation}
\Phi(r)=\Phi_{B}(r)+\delta \Phi_{\nu}(r),
\label{eq:115}
\end{equation}
where $\Phi_{B}(r)$ is the Newtonian potential due to the sun
($M_{B}=M_{\odot}$), i.e.
\begin{equation}
\label{eq:116}
 \Phi_{B}(r)=-\frac{GM_{\odot}}{r},
\end{equation}
and $\delta \Phi_{\nu}(r)$ is the contribution 
 of the neutrino halo  to the potential
\begin{equation}
\delta\Phi_{\nu}(r)=\frac{GM_{\odot}}{a_{\nu}}\left(\frac{1-v(x)}{x}+v'(x_{0})
\right).
\label{eq:117}
\end{equation}
Here $v$ is the  potential variable  that satisfies the 
Lan\'e-Emden equation
with  a pointlike solar mass source at the center, i.e. a solution
of the equations (\ref{eq:111}) and (\ref{eq:112}) for $M_{B}=M_{\odot}$.

In the vicinity of the sun, the  term
$\delta \Phi_{\nu}(r)$ is much smaller than $\Phi_{B}(r)$. 
Near the center, the
potential  variable $v(x)$ has the asymptotic behaviour
\begin{equation}
v(x) \approx \frac{-4}{3}x^{3/2}+v'(0)x+1 \ ,
\label{eq:118}
\end{equation}
where $v'(0)$ 
 parametrizes  solutions with different halo  masses.
Inserting the last expression
into Eq.~(\ref{eq:117}),
we find for the additional potential energy
  due to the neutrinos
\begin{equation}
\delta U=m \delta \Phi_{\nu}(r)=\frac{4mGM_{\odot}}{3a_{\nu}^{3/2}}r^{1/2}+C,
\label{eq:119}
\end{equation}
where $C =v'(x_{0})-v'(0)$ is a constant which is irrelevant for our
further
considerations, and $m$ is the mass of the planet or asteroid. Thus
our problem reduces to  investigating
the perihelion shifts of the planets and 
asteroids due to the  small perturbation given
  by Eq.~(\ref{eq:119}).

When a small correction $\delta U(r)$ is added to the potential
energy $U=-\alpha/r$, the paths of finite motion are no longer closed,
and at each revolution, the perihelion is displaced
by a small angle  (Landau and Lifshitz 1960) 
\begin{equation}
\delta \varphi=\frac{\partial}{\partial L} \left(\frac{2m}{L}
\int_{0}^{\pi}r^{2} \delta U d\varphi \right).
\label{eq:120}
\end{equation}
Here $L$ is the angular momentum of the planet or asteroid whose 
 unperturbed orbit   is given by the standard
 equation for an ellipse, i.e.
\begin{equation}
\label{eq:121}
r=\frac{p}{1+e\cos \varphi},
\end{equation}
with
\begin{equation}
\label{eq:122}
 p=a(1-e^{2}), \    e=1+\frac{2EL^{2}}{m\alpha^{2}}, \
a=\frac{\alpha}{2E}, \  \alpha=mGM_{\odot}, \  L^{2}=pm \alpha,
\end{equation}
where $E$ is the energy of
 the planet or asteroid $ (E < 0)$, $e$ is the eccentricity,
$a$  the  semi-major  axis and $p$  the latus rectum of its orbit.

Before calculating the perihelion shifts due to the  neutrino
halo, let us consider a general perturbation potential 
of the power law form
\begin{equation}
\label{eq:123}
\delta U=\gamma r^{n},
\end{equation}
where $\gamma$  and $n$ are constant  real numbers.
For instance, it
 is well known that
the general relativistic  corrections for  a planet
moving around the sun can be incorporated in the Newtonian
framework  via  an effective
potential of the form 
\begin{equation}
\label{eq:124}
U_{eff}=-\frac{\alpha}{r}+\frac{L^{2}}{2mr^{2}}-\frac{\beta}{r^{3}},
\end{equation}
where the constant $\beta$ is 
given by
\begin{equation}
\label{eq:125}
\beta=\frac{GM_{\odot}L^{2}}{mc^{2}} .
\end{equation}
The last term in the Eq.~(\ref{eq:124}), is 
the  general relativistic correction
to  the potential $\delta U$ with $\gamma=\beta$ and $n=-3$.
 Due to this term, the perihelion
is shifted by
an angle $\delta \varphi_{GRT}$ which can be easily calculated from  
 Eq.~(\ref{eq:120}) yielding
\begin{eqnarray}
\label{eq:126}
\delta \varphi_{GRT}&=&\frac{\partial}{\partial L}\bigg[\frac{2m}{L}
\int_{0}^{\pi}r^{2}\frac{-\beta}{r^{3}}d\varphi\bigg]=-2m^{2}\alpha\beta
\frac{\partial}{\partial L}
\bigg[\frac{1}{L^{3}}\int_{0}^{\pi}(1+e\cos\varphi)d\varphi\bigg]
 \nonumber\\
&=&6\pi m^{2}\beta\alpha L^{-4}=\frac{6\pi GM_{\odot}}{c^{2}a(1-e^{2})} 
\end{eqnarray}
  for the general
relativistic perihelion shift per revolution.
Let us assume that,
in addition to the general relativistic perturbation potential $-\beta/r^{3}$,
 there is a further perturbative potential given
 by Eq.~(\ref{eq:123}) due to
some type of DM.
It is clear that this small shift will be additive  in first order.
The contribution to  the shift per revolution  due to the perturbation
potential $\delta U$ is thus  
\begin{equation}
\label{eq:127}
\delta\varphi=\frac{\partial}{\partial L}\bigg[\frac{2m}{L}\int_{0}^{\pi}
\gamma r^{n}r^{2}d\varphi\bigg]=2m\gamma\frac{\partial}{\partial L}\bigg[\frac{1}{L}
\int_{0}^{\pi}\frac{p^{n+2}}{\left(1+e\cos\varphi\right)^{n+2}}d\varphi\bigg]
.
\end{equation} 
After some algebraic manipulations,
we arrive at the  expression for the shifts per revolution
\begin{equation}
\label{eq:128}
\delta \varphi =\frac{2 \gamma}{\alpha}\bigg(a(1-e^{2})\bigg)^{n+1}
\bigg[(2n+3)I_{n+2}(e)+
\frac{1-e^{2}}{e^{2}}(n+2)\bigg(I_{n+2}(e)-I_{n+3}(e)\bigg)\bigg] ,
\end{equation}
where the function $I_{n}(e)$ has been defined as
\begin{equation}
\label{eq:129}
I_{n}(e)=\int_{0}^{\pi}\frac{1}{(1+e\cos\varphi)^{n}}d\varphi
=\frac{\pi}{\left(1-e^{2}\right)^{n/2}}
P_{n-1}\bigg(\frac{1}{\sqrt{1-e^{2}}}\bigg) ,
\end{equation}
$P_{n-1}(x)$ being  the Legendre  polynomial of degree $n-1$.
In the limit of small eccentricities, i.e. $e \rightarrow 0$,
 Eq.~(\ref{eq:128}) yields 
\begin{equation}
\label{eq:130}
\delta\varphi =-\frac{\gamma \pi}{\alpha} a^{n+1}n(n+1) .
\end{equation}
The perihelion shifts become  negative, i.e.
$\delta\varphi < 0$ for $n > 0$ and $n < -1$ assuming
 of course $\gamma > 0$.
In the following, we will deal with two
special cases: degenerate  neutrino DM  and
DM with a constant density which could describe homogeneous WIMP DM  in the
vicinity of the sun.

In the case of a neutrino halo,
the perturbative potential $\delta U$ is given by
  Eq.~(\ref{eq:119}). We can 
  thus apply  Eq.~(\ref{eq:128}),
with  $n=1/2$ and  $\gamma=4mGM_{\odot}/(3a_{\nu}^{3/2})$.
 The constant $C$ in Eq.~(\ref{eq:119}) does not contribute
 to the perihelion shift.
After integration, we arrive at the perihelion shift per revolution 
\begin{equation}
\label{eq:131}
\delta \varphi=-\frac{8}{3}\left(\frac{a}{a_{\nu}}\right)^{3/2}
\frac{(e+1)(1-e)^{1/2}}{e^{2}}
\bigg[E(\pi/2,k)+(e-1)F(\pi/2,k)\bigg],
\end{equation}
where $k$ is given by
\begin{equation}
k=\sqrt{\frac{2e}{1+e}} \ ,
\label{eq:132}
\end{equation}
and $F(\pi/2,k)$ and $E(\pi/2,k)$ are the complete  elliptic integrals
of the first and second kind, respectively, defined as
\begin{equation}
F(\pi/2,k)=\int_{0}^{\pi/2} \frac{d\varphi}{\sqrt{1-k^{2}\sin^{2}\varphi}} \ ,
\label{eq:133}
\end{equation}
\begin{equation}
E(\pi/2,k)=\int_{0}^{\pi/2} d\varphi \sqrt{1-k^{2}\sin^{2} \varphi} \ .
\label{eq:134}
\end{equation}
We can now  write the expression for
the perihelion shift per century as
\begin{eqnarray}
\delta\tilde{\varphi} & =& -5.50''\times10^{5}\left(\frac{a_{E}}{a_{\nu}}\right)^{3/2}
\frac{(e+1)(1-e)^{1/2}}{e^{2}} \nonumber \\
 & &\times \bigg[E(\pi/2,k)+(e-1)F(\pi/2,k)\bigg]
\times \frac{100}{T_{E}} ,
\label{eq:135}
\end{eqnarray}
where   Kepler's third law has been used to  eliminate  
the period $T$ of the planet, i.e. 
\begin{equation}
T= T_{E}\frac{a^{3/2}}{a_{E}^{3/2}}.
\label{eq:136}
\end{equation}
Here, $T_{E}$ and $a_{E}$ are the  period and the semi-major
axis of the Earth's orbit, respectively.
From Eq.~(\ref{eq:135}), we 
 conclude that the  shift does  not depend on the semi-major axis $a$, i.e.
they 
are a function of the eccentricity $e$ only. Moreover, the shift is  
 negative in contrast to the general 
relativistic one 
(see Eq.~(\ref{eq:126})).
In Fig.~5, we present the shift due a neutrino halo as
a function of the eccentricity
for various neutrino masses.
 The data points show
the difference between the 
observed perihelion shift (obs) and the correction due to the general
relativity theory (GRT), i.e.
 $\delta \tilde{\varphi}_{obs}-\delta\tilde\varphi_{GRT}$.
Thus,
if there was no neutrino halo, $\delta \tilde{\varphi}$ should
vanish identically.
The general relativistic corrections, calculated
using the exact general relativistic
equations rather than the useful approximation
eq.~( \ref{eq:135}), as well as the observed values are
taken from Weinberg (1972).
The data points with error bars are shown for Venus, Earth, Mercury and 
Icarus.
As expected,
the error bars of the data points decrease for increasing eccentricity.
We may thus obtain a good
 upper bound for the neutrino mass
using  the perihelion shift data for  Icarus.
In Fig.~6, the expected neutrino halo  shift 
 is  plotted  for Icarus as a function
of the neutrino mass. The horizontal lines denote
the region of
 $\delta \tilde{\varphi}=\delta \tilde{\varphi}_{obs}-\delta\tilde\varphi_{GRT}$
allowed by observations.
The observational shifts for Icarus is
$\delta \tilde{\varphi}_{obs}=9.8''\pm 0.8''$ per century, while the general
relativistic correction is $\delta \tilde{\varphi}_{GRT}=10.3''$
per century (Weinberg 1972). Requiring that the shift due to
a possible neutrino halo
cannot be 
smaller than  the lower limit of 
$\delta \tilde{\varphi}=\delta \tilde{\varphi}_{obs}-\delta \tilde{\varphi}_{GRT}=-0.5''\pm 0.8''$,
i.e. $\delta \tilde{\varphi} \stackrel{\textstyle >}{\sim} -1.3''$ per
century,
we obtain
an  upper limit  for the  neutrino mass
of
\begin{eqnarray}
&m_{\nu}c^{2}& \leq 16.4 \ {\rm keV} \   {\rm for} \  g_{\nu}=2, \nonumber\\
&m_{\nu}c^{2}& \leq  13.8 \ {\rm keV} \  {\rm for} \  g_{\nu} =4 \ ,
\label{eq:137}
\end{eqnarray}
as seen from Fig.~6. Here, we note that the 
bounds (\ref{eq:137}) agree very well with those obtained
from the mass excesses
 in Figs.~1 and 2. 

Moreover, considering the violent supermassive dark object
at the center of M87, with  mass $M=(3.2 \pm0.9) \times 10^{9}M_{\odot}$
(Macchetto et al. 1997), as
a relativistic neutrino ball at the Oppenheimer- Volkoff limit (Bili\'c, Munyaneza
and Viollier 1999), the neutrino mass
is constrained by
\begin{eqnarray}
&m_{\nu}c^{2}& \leq 16.5~{\rm keV}  \ {\rm for} \ g_{\nu}=2 \nonumber\\
&m_{\nu}c^{2}& \leq  13.9~{\rm keV} \ {\rm for} \ g_{\nu}=4  \ .
\label{eq:137a}
\end{eqnarray}
Such a neutrino ball 
would be virtually indistinguishable from a supermassive black hole,
as its radius is 4.45 Schwarzschild radii, little more than the radius
of the last stable orbit around a black hole of three Schwarzschild radii.
In fact, as the mass density is very small
near the surface of the neutrino ball, the ''effective'' neutrino
ball radius is substantially smaller than 4.45 Schwarzschild radii.
Furthermore, if we interpret  the supermassive dark object
of mass $M=(2.6 \pm 0.2)\times 10^{6}M_{\odot}$
at the Galactic center (Ghez et al. 1998)  as 
a neutrino ball (Viollier, Trautmann and Tupper 1993),
the observed motion of stars close to Sgr A*  yields
an upper limit for the radius of the compact dark object, and therefore
 a lower 
limit for the neutrino mass (Munyaneza, Tsiklauri and Viollier 1998, 1999).
Finally,
the infrared drop of the emission spectrum
of Sgr A*, interpreted in terms of
standard thin accretion disk theory, provides us with an upper limit
of the neutrino mass
(Bili\'c, Tsiklauri and Viollier 1998, 
Tsiklauri and Viollier 1999; Munyaneza and Viollier 1999),
because of the cutoff of the emission of disk
radiation inside the neutrino ball. Combining  these two constraints,
we obtain
\begin{eqnarray}
15.9 \ {\rm keV} \leq &m_{\nu}c^{2}& \leq 25 \ {\rm keV} \ {\rm for} \
g_{\nu}=2,
\nonumber \\
13.4 \ {\rm keV} \leq &m_{\nu}c^{2}& \leq 21\ {\rm keV} \  {\rm for} \
g_{\nu}=4.
\label{eq:137b}
\end{eqnarray}
Such a neutrino ball would differ substantially from
a supermassive black hole, as the escape velocity from the center
would be only about $v_{\infty} \approx 1700~ {\rm km/s}$
and the radius is about $4\times 10^{4}$ Schwarzschild radii.
Virtually all supermassive compact dark objects
that have been   observed so far at the centers of galaxies have
masses in  the range of $10^{6.5}$ to $10^{9.5}M_{\odot}$ and could
therefore be explained in the neutrino ball scenario.
 
We now  turn to the interpretation of DM in the solar system 
at  constant density (Gr\o n and Soleng 1996) which could describe
WIMP DM that is not clustered around the sun.
In this case, we have  $n=2$ and  $\gamma = 2\pi mG\rho_{d}/3$,
where $\rho_{d}$ is the density of DM in the solar system.
Eq.~(\ref{eq:128}) yields  for the shifts per century 
\begin{equation}
\label{eq:139}
\delta\tilde{\varphi}=-\frac{4\pi^{2}\rho_{d}a^{3}}{M_{\odot}}(1-e^2)^{1/2}
\times \frac{100}{T} .
\end{equation}
Here $T$ is the  Icarus period   and $a$  its semi-major axis .
From Eq.~(\ref{eq:139}), we gather that 
 the shifts are negative, as in the previous  case,
 but they now  depend on two parameters: the semi-major
axis $a$ and the eccentricity $e$.
We can establish an upper limit on the
 density of DM using the  Icarus data.
 Its eccentricity  is $e= 0.827$,
the semi-major axis  
$a=1.076~{\rm AU}$ and the period $T=1.116~{\rm yr}$ .    
By requiring that the DM  
shift  is restricted to 
 $\delta \tilde{\varphi} \stackrel {\textstyle >}{\sim} -1.3''$ per
century,
we get
 an upper  bound on  the
DM density in the solar system of 
\begin{equation}
\label{eq:141}
\rho_{d} \stackrel {\textstyle < }{\sim} 1.5\times 10^{-15}~{\rm g/cm}^{3}.
\end{equation}
Such a  constant density DM  model was studied 
by Gr\o n and Soleng (1996) in the framework
of general relativity.
However, they derived an upper limit of
the DM density of
 $\rho_{d} \stackrel {\textstyle <}{\sim} 1.8 \times 10^{-16} {\rm g/cm^{3}}$ 
from the perihelion motion of Icarus.
If we use $-0.8''$
instead of $-1.3''$
for the lower limit of the DM shift of Icarus, as
Gr\o n 
and Soleng (1996) did,
we obtain a density 
of $\rho_{d} \stackrel {\textstyle <}{\sim} 9.4 \times 10^{-16} {\rm g/cm^{3}}$
which  is still larger by a factor of $5.2$ than the upper limit obtained by 
Gr\o n and Soleng (1996).
The discrepancy between our results  and  those by Gr\o n and Soleng
can be traced back
 to the fact that those authors used for their calculations a
 perturbation expansion in terms of the eccentricity valid only for nearly
circular orbits ($e \approx 0$) while Icarus has an eccentricity of $e=0.827$.
Using our  value of  the upper  limit for $\rho_{d}$, i.e.  Eq.~(\ref{eq:141}),
one can calculate the total  DM mass
within the orbits
of the various planets.
Within the Earth's orbit, the DM mass is 
$M_{d} \leq 1.1\times10^{-8}M_{\odot}$ which is in agreement
with the bound of
 $M_{d} \stackrel {\textstyle <}{\sim} 7 \times 10^{-8}M_{\odot}$
 obtained by Mikkelsen and Newman (1977).
  Within Uranus' orbit , we get a DM mass bound
 $M_{d} \leq 7.6 \times 10^{-5}M_{\odot}$, and, finally for the DM 
 mass within
Neptune's orbit we have  $M_{d} \le 2.9\times10^{-4}M_{\odot}$.
 The last two bounds  are clearly in conflict with the observed
ephemeris, which allow only a  DM  mass of  the order of a few times
$10^{-6}M_{\odot}$, even
 within the orbit of Neptune (Anderson et al. 1995).
We thus conclude  that if the DM density is constant, the upper bound
of DM within Neptune's orbit restricts the DM density in the solar system to
 $\rho_{d} \stackrel {\textstyle <}{\sim} 1.5 \times 10^{-17} {\rm
g/cm^{3}}$.

\section{Other observational consequences of a degenerate neutrino halo}

We now turn to the question whether
a neutrino halo, with properties as described in the
last sections, can be observed in nature, using other than
gravitational detection techniques.
Let us consider the most conservative scenario, in  which the
Standard Model of Particle Physics (SMP) , minimally modified
to accommodate three species
of massive neutrinos, that are mixed
through a leptonic Cabibbo-Kobayashi-Maskawa (CKM) matrix, is basically
correct at low energies.
We further assume that  our heavy neutrino
 is a Dirac neutrino, more specifically the  mass eigenstate $\nu_{\tau}$
, in the mass range between $10~{\rm keV}/c^{2}$ and $25~ {\rm keV}/c^{2}$,
while the $\nu_{e}$ and $\nu_{\mu}$ are assumed to be massless.
The $\nu_{\tau}$ couples preferentially to the $\nu'_{\tau}$ and
to a lesser extent   to the $\nu'_{e}$  and $\nu'_{\mu}$ eigenstates of the charged weak
interaction.
In the framework of the  SMP, the dominant decay mode of the mass
eigenstate $\nu_{\tau}$ in the assumed mass range is
 the conventional radiative decay, having  a lifetime
(Boehm and Vogel 1987) of
\begin{equation}
\tau_{D}(\nu_{\tau} \rightarrow \nu_{i}\gamma)=1.30 \times 10^{15}
\left(\frac{17.2~{\rm keV}}{m_{\nu}c^{2}}\right)^{5}
|U_{\tau\nu_{\tau}}U^{*}_{\tau\nu_{i}}|^{-2} \  {\rm yr}. 
\end{equation}
Here the CKM- matrix element $U_{\tau\nu_{i}} (i=e,\mu)$
with the largest modulus, presumably $i=\mu$, determines the dominant decay mode.
The present limit from the CHORUS collaboration for 
$|U_{\tau\nu_{\mu}}|^{2}$
is $|U_{\tau\nu_{\mu}}|^{2} < 3.3\times 10^{-4}$ for
$\delta m_{\tau\mu}^{2} \stackrel {\textstyle >}{\sim} 100~{\rm eV}^{2}$
(Sato 1999).
One can thus safely
conclude that the $\nu_{\tau}$ is quasistable
over the lifetime of the Universe.
However, even though the $\tau$-neutrino
is remarkably stable
against radiative decay,
$\tau$-neutrino matter is quite radioactive. It is  perhaps so abundant
that  the X-ray
flux generated by the $\tau$-neutrino decay rates
\begin{equation}
\dot{n}_{\nu}=-\frac{1}{\tau_{D}}n_{\nu},
\end{equation}
where the neutrino (and antineutrino) number density $n_{\nu}$ is 
obtained from 
Eq.~(\ref{eq:112a})
\begin{equation}
n_{\nu}(r)=\frac{\rho_{\nu}(r)}{m_{\nu}}=
\frac{8m_{\nu}^{7}g_{\nu}^{2}G^{3}M_{\odot}^{2}}
{9\pi^{3}\hbar^{6}}\bigg(\frac{v}{x}\bigg)^{3/2},
\end{equation}
could be observable.
 In the vicinity of a pointlike baryonic star of mass
$M_{B}$, i.e. $v \approx M_{B}/M_{\odot}$, may be rewritten as
\begin{equation}
n_{\nu} \approx \frac{m_{\nu}^{3}g_{\nu}}{6\pi^{2}\hbar^{3}}
\left(\frac{2GM_{B}}{r}\right)^{3/2}
=0.978 \times 10^{17} g_{\nu}
\left(\frac{M_{B}R_{\odot}}{M_{\odot}r}\right)^{3/2}
\left(\frac{m_{\nu}c^{2}}{17.2~{\rm keV}}\right)^{3}
 {\rm cm}^{-3} .
\end{equation}
Thus the number of photons with energy $m_{\nu}c^{2}/2$  emitted per unit
time    
and volume is
\begin{equation}
\dot{n}_{\nu}=-75.2 g_{\nu}
\left(\frac{M_{B}R_{\odot}}{M_{\odot}r}\right)^{3/2}
\left(\frac{m_{\nu}c^{2}}{17.2~{\rm keV}}\right)^{8}
|U_{\tau\nu_{\tau}}U^{*}_{\tau\nu_{i}}|^{2} {\rm cm}^{-3} {\rm yr}^{-1},
\end{equation}

Although an energy of $m_{\nu}c^{2}/2 \approx  8~{\rm keV}$ is equivalent
to a temperature of
roughly $10^{8}$ K, the X-ray flux from neutrino decays near the solar
surface is too small
to contribute
significantly to  nuclear synthesis in the sun or to maintaining
the solar corona at a temperature of a few million degrees.
At a distance of one astronomical unit ($r=a_{E}$), the number density
and decay rates per unit volume are
\begin{equation}
n_{\nu} \approx 3.10 \times 10^{13} g_{\nu}
 \bigg(\frac{m_{\nu}c^{2}}{17.2 \ {\rm keV}}\bigg)^{3} {\rm cm}^{-3},
\end{equation}
and
\begin{equation}
\dot{n}_{\nu} \approx -2.38\times 10^{-2} g_{\nu}
\left(\frac{m_{\nu}c^{2}}{17.2~{\rm keV}}\right)^{8}
|U_{\tau\nu_{\tau}}U^{*}_{\tau\nu_{\mu}}|^{2} {\rm cm}^{-3} {\rm yr}^{-1},
\end{equation}

Thus, if our model of the solar neutrino halo is correct, we  predict
that in a shielded vacuum of  $1000~{\rm m}^{3}$, one will observe
$g_{\nu}$ photons per hour for
$|U_{\tau\nu_{\tau}}U^{*}_{\tau\nu_{\mu}}|^{2}=3.3\times 10^{-4}$
which is the present experimental  limit of the CHORUS collaboration,
 and $m_{\nu}= 16~{\rm keV}/c^{2}$, $g_{\nu}$ being
the spin degeneracy factor of
 neutrinos (Viollier, Leimgruber and Trautmann 1992).
However , this number could be substantially
enhanced  through the gravitational field of the Earth.
 The photons
originating from the the radiative
 decay
 $\nu_{\tau} \rightarrow \nu_{\mu}+\gamma $ 
or ( $\nu_{\tau} \rightarrow \nu_{e}+\gamma$)
will have a sharp energy of $m_{\nu}c^{2}/2$ with 
$\Delta E/E \approx  10^{-4}$.
This measurement could prove
the existence of the massive neutrino halo and fix the $\nu_{\tau}$
mass and the $\nu_{\tau}-\nu_{\mu}$  ( or $\nu_{\tau}-\nu_{e}$
mixing angle accurately. At the same time, this could be the first
direct evidence for the neutrino background and the nature of DM.

If both  neutrinos and antineutrinos are
present in a neutrino halo around the sun, these will annihilate
into light  neutrinos $\nu_{e}$, $\overline{\nu}_{e}$ and
$\nu_{\mu}$, $\overline{\nu}_{\mu}$ through ordinary
weak interactions  processes via the $Z^0$,  which are independent
of the mixing angle of course. In fact, the rate  of change
of the neutrino (and antineutrino) number density is
\begin{equation}
\dot{n}_{\nu}=-<\sigma_{A}\upsilon_{\nu}>\frac{n_{\nu}^{2}}{2},
\end{equation}
where the spin averaged annihilation  cross section
is for Dirac neutrinos given by
\begin{equation}
<\sigma_{A}\upsilon_{\nu}>=\frac{G_{F}^{2}m_{\nu}^{2}c}{\pi\hbar^{4}g_{\nu}} .
\end{equation}
The largest annihilation rate is obtained  in the interior of the star
 which
of course depends on the internal structure
 of the star.
 However, the effective
annihilation time  $\tau_{A}$ at the surface
of the star with mass $M_{B}$ and radius $R_{B}$  is a good
indicator for how fast this process actually is.
It can be calculated quite reliably yielding
\begin{eqnarray}
\tau_{A}&=&\bigg|\frac{n_{\nu}}{\dot{n}_{\nu}}\bigg|=
\frac{2}{<\sigma_{A}\upsilon_{\nu}>n_{\nu}}
 \approx \frac{12\pi^{3}\hbar^{7}}{G_{F}^{2}m_{\nu}^{5}c}
\left(\frac{R_{B}}{2GM_{B}}\right)^{3/2}
\nonumber\\ 
&=&
0.4335\times 10^{13}\left(\frac{17.2~{\rm keV}}{m_{\nu}c^{2}}\right)^{5}
\left(\frac{R_{\odot}M_{\odot}}{R_{B}M_{B}}\right)^{3/2} {\rm yr}.
\end{eqnarray}
Thus for a neutrino mass of $m_{\nu}=16~{\rm keV}/c^{2}$
and a solar mass $M_{B}=M_{\odot}$ with a radius 
$R_{B}=R_{\odot}$,
we obtain $\tau_{A}=0.623\times 10^{13}{\rm yr}$, much larger than  the
age of the universe. Although this annihilation process is more
efficient than the radiative decay, it will be essentially
unobservable due to the low energy of the neutrinos.

\section{Summary}
In this paper, we have investigated
the properties and implications  of  a possible halo of degenerate
 neutrino  matter
around the sun.
For small halo masses or sufficiently close to the center, this
neutrino halo is dominated by   the gravitational potential of
the sun.
We have established that the enclosed mass of a degenerate neutrino halo
around the sun is, for neutrino masses of $m_{\nu} \sim 15 ~{\rm keV}/c^{2}$, 
of the order of   a few  times $10^{-6}M_{\odot}$ within 
 Uranus' and Neptune's orbits, 
 consistent with available observational data.
If  such  a neutrino halo
exists, it would decrease the perihelion shifts, i.e.
the  neutrino halo shifts   are negative in contrast
to the general relativistic ones.
The perihelion shifts  due  to the neutrino halo
depend only on one parameter, 
the eccentricity $e$, while those due to general relativistic effects
depend on the eccentricity $e$ and the semi-major axis $a$.
The maximal radius of such a degenerate neutrino 
halo around the sun is 
 a few light years, with a total halo mass of
 $\sim 3~M_{\odot}$.

In order to explain the mass excesses
within  the orbits of various outer planets using the Voyager 1/2
 and Pioneer 10/11 data, the neutrino
mass should be in a narrow range of $m_{\nu}= (15\pm 1)~{\rm  keV}/c^{2}$
for $g_{\nu}=2$.
The predicted values of the acceleration in the solar system
 have been compared to those obtained from
 the recent observations.
We have seen that a degenerate heavy neutrino halo
around the sun fits the recent  ephemeris very  well,
but it cannot explain  the anomalous
 acceleration  in the Pioneer data (Anderson et al. 1998; Turyshev 1999).
 We  have shown that, in order 
 to be consistent with the observations,
 the total mass of the neutrino halo
should be less than   $\sim 10^{2}M_{\odot}$.
A neutrino mass in the range from 14 to $16~{\rm keV}/c^2$ fits 
all the observational data on  DM in  the solar
system, the compact dark central object Sgr A$^{*}$ in our Galaxy
with $M=(2.6\pm 0.2)\times 10^{6}M_{\odot}$, and the
most massive  compact dark object at the center of the  galaxy M87 with
$M=(3.2\pm0.9)\times 10^{9}M_{\odot}$.

We have established 
an upper bound of DM
with  a constant  density  around the sun
of $\rho_{d} \stackrel {\textstyle <}{\sim} 1.5\times 10^{-17}{\rm g/cm^{3}}$
using the upper limit for the DM mass within Neptune's orbit.

We have proposed a new  experiment aimed at observing
the radiative decay of the neutrinos in the mass range around
$15~{\rm keV}/c^{2}$,  which the sun might have accumulated
in a degenerate neutrino halo during its formation.

This research was supported by the Foundation for Fundamental Research.
F.~Munyaneza  gratefully acknowledges 
funding from Deutscher Akademischer Austauschdienst (DAAD). 
We wish also to thank R.~Lindebaum for useful discussions.

\newpage
\centerline{Figure captions:}

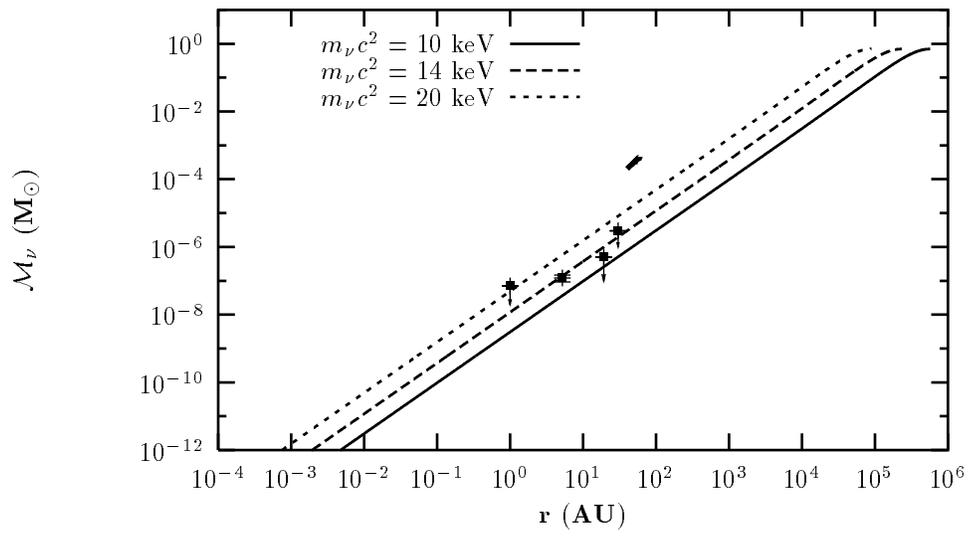
Fig1: Mass ${\cal M}_{\nu}(r)$ of  degenerate neutrino matter
  enclosed within a radius $r$
 from
the sun for various neutrino masses. The total mass of the halo
 is $0.7 M_{\odot}$.
 The observational data points with error bars
(Anderson et al. 1995) are shown for Jupiter ($a=5.2~ {\rm AU}$),
Uranus ($a=19.2~{\rm AU}$) and Neptune ($a=30.1~ {\rm AU}$).
The data point at $r=1~{\rm AU}$  is taken from Mikkelsen and Newman (1977).
 The arrows indicate upper limits.
The dark mass needed to explain the anomalous acceleration (Anderson et al.
1998; Turyshev et al. 1999)
is indicated by  a bar at 40 to $60~{\rm AU}$.

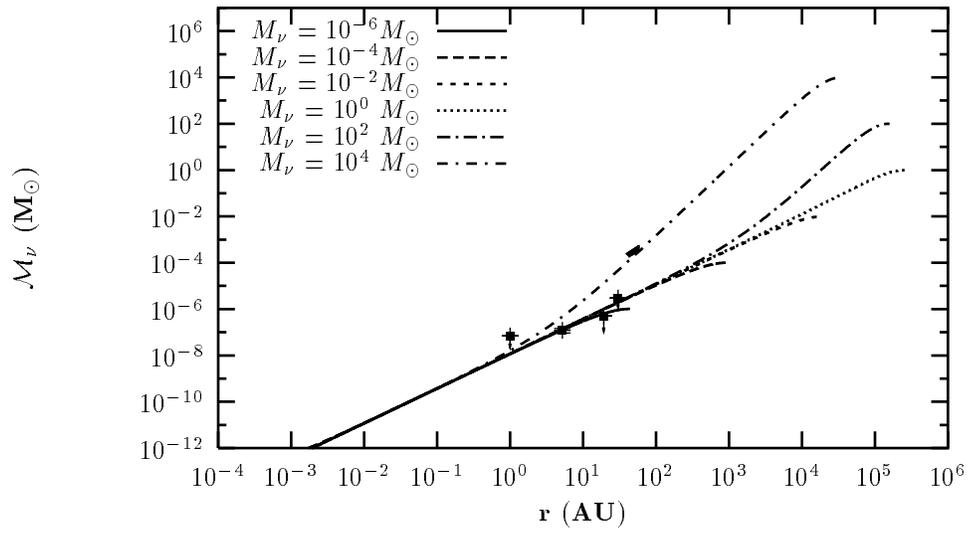
Fig2: Mass ${\cal M}_{\nu}(r)$ of degenerate neutrino matter enclosed
 within a radius $r$ from the sun for various total 
masses $M_{\nu}$ of the neutrino halo.
The neutrino mass  and degeneracy factor are  fixed to
$m_{\nu}=14~{\rm keV}/c^2$ and 
$g_{\nu}=2$, respectively. A total mass of the neutrino halo which is less
than  $M_{\nu} \sim 100 M_{\odot}$ would 
be consistent with the observed mass excess data
from Pioneer 10 and 11 (Anderson et al. 1995). The data points
with error bars are shown for Jupiter, Uranus and Neptune. 
The observational data at $r=1 {\rm AU}$ is taken from
Mikkelsen and Newman (1977). The bar at 40 to $60~{\rm AU}$ represents
the dark mass needed to explain the anomalous acceleration of
 $a_{P}=7.5\times 10^{-8}{\rm cm/s}^{2}$.

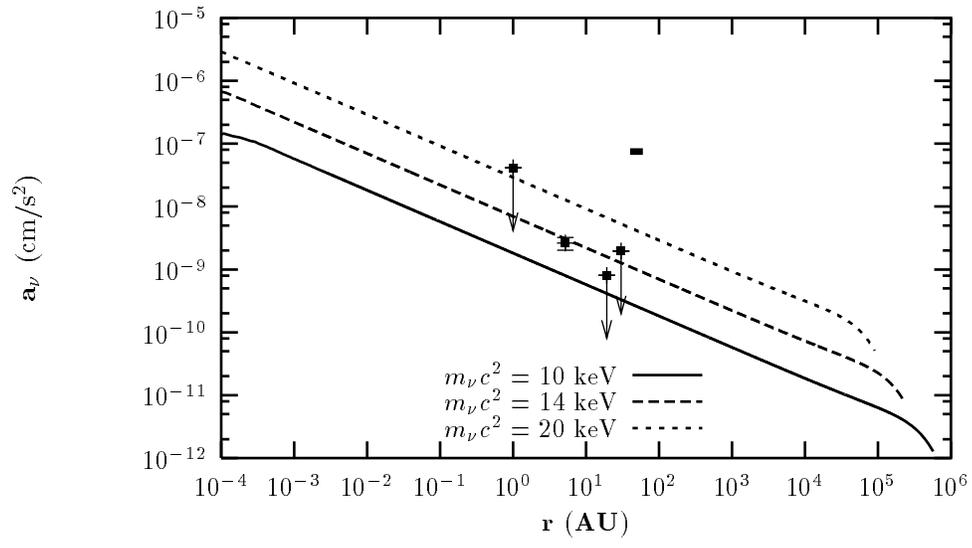
Fig3: Excess acceleration $a_{\nu}$ due to a neutrino halo
around the sun. The total mass of the halo is $M_{\nu}=0.7 ~M_{\odot}$.
The expected values of the acceleration using observational
data points (Anderson et al. 1995)
 are shown by points with arrows. The point with arrow 
at $r=1~{\rm AU}$ is calculated
using the data from Mikkelsen and Newman (1977).
The anomalous acceleration at 40 to 60 AU is indicated by a horizontal bar.

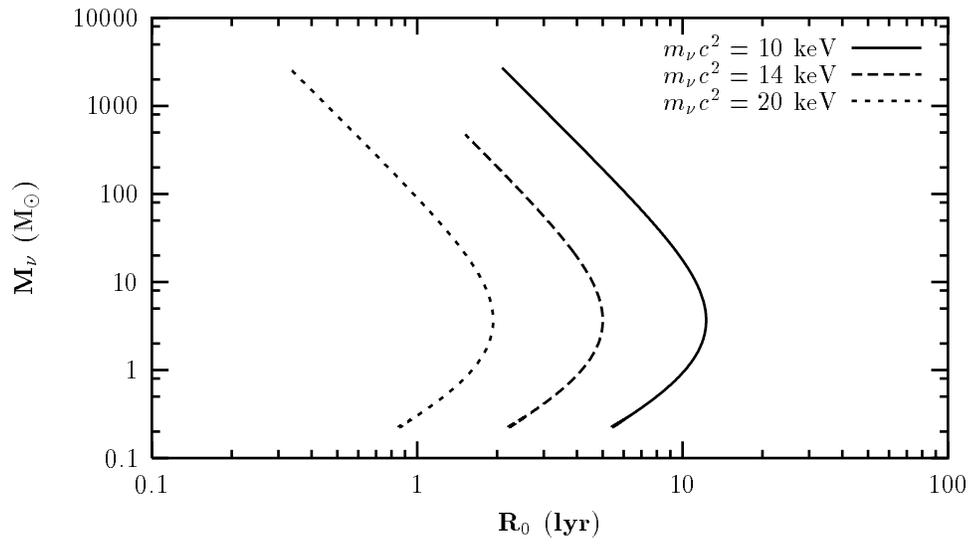
Fig4: The total  mass $M_{\nu}$  as a function
 of the radius $R_{0}$ of the neutrino halo
around the sun.
 The neutrino mass $m_{\nu}$ is varied as shown on the graph. 
$M_{\nu,max}$ turns out to be fairly constant, 
i.e. $M_{\nu,max} \approx 3 M_{\odot}$,
for the maximal radius $R_{max}$ of the neutrino halo.

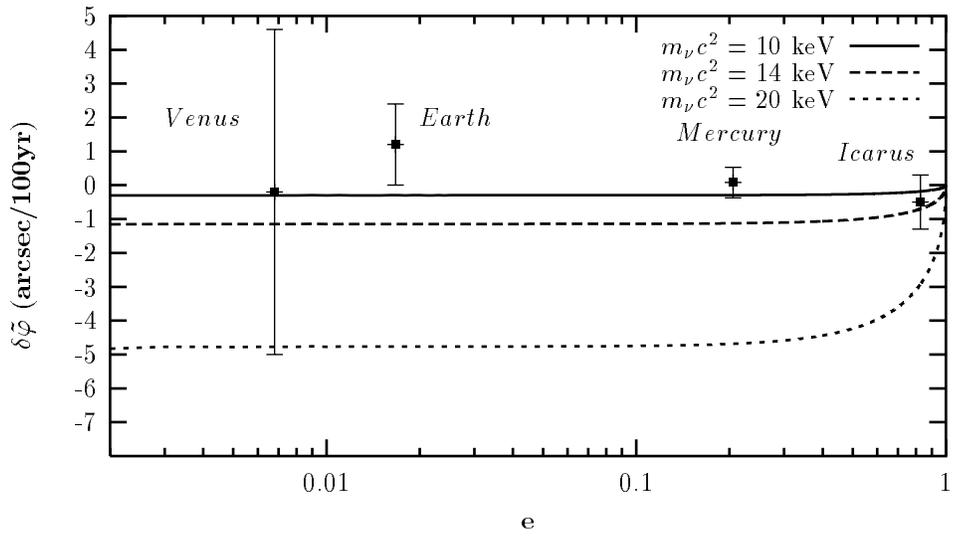
Fig5: Perihelion shifts $\delta\tilde{\varphi}$ caused by a neutrino halo
as a function of the eccentricity $e$. The data points with
error bars denote the difference between the observed perihelion shifts
 and
 general relativistic  corrections for the perihelion shifts.
The data points shown are for Venus ($e=0.007$), Earth ($e=0.017$),
Mercury ($e=0.206$) and Icarus ($e=0.827$). 

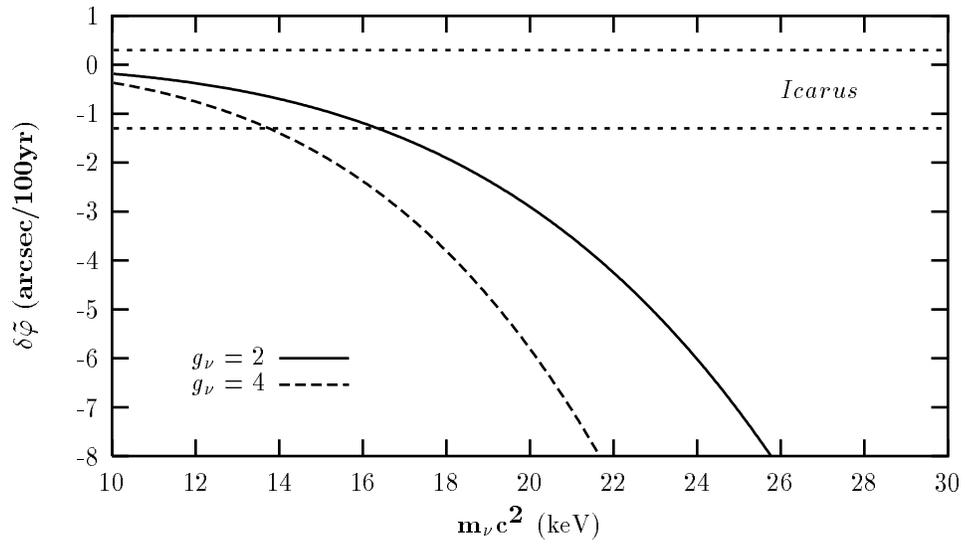
Fig6: The Icarus perihelion shift $\delta \tilde{\varphi}$ as a function 
of the neutrino mass $m_{\nu}$.
The horizontal lines show the difference between the observed  value 
 and the correction  predicted by  general  relativity  theory .
In order to be consistent with the observational data for the Icarus
perihelion shift, the neutrino mass is constrained by
$m_{\nu}c^{2} \le 16.4~{\rm keV}$ for $g_{\nu}=2$ or
 $m_{\nu}c^{2} \le 13.8~{\rm keV}$ for $g_{\nu}=4$.


\begin{references}

\reference{}
Alcock , C. {\it et al.} 1996, ApJ, 461, 81
\reference{}
Alcock, C.  {\it et al.} 1997, ApJ, 486, 697 
\reference{}
Anderson, J. D., Lau,  E. L., Krisher, T. P.,Dicus , D. A.,
Rosenbaum, D. C., and  Teplitz, V. Z.  1995, ApJ, 448, 885;
 1996, ApJ, 464, 1054(E)
\reference{}
Anderson, J. D. , Laing,P.A., Lau, E. L., Liu, A. S., Nieto, M.M., and
Turyshev S. G. 1998, Phys. Rev. Lett.,  81, 2858
\reference{}
Ansari, R. {\it et al.} 1996, A\&A, 314, 94
\reference{}
Ashman, K. M. 1992, PASP, 104, 1109
\reference{}
Bahcall, J. N. 1984a, ApJ, 276, 169
\reference{}
Bahcall, J. N. 1984b, ApJ, 287, 926
\reference{}
Bekenstein, J. D.  1992,  in
 Proceedings of the 6th Marcel Grossmann Meeting on General Relativity
 ed. H. Sato, and T. Nakamura (World Scientific, Singapore), 905
\reference{}
Bili\'c, N. , Munyaneza, F., and Viollier, R. D.  1999, Phys. Rev., D59, 024003
\reference{}
Bili\'c, N. , Tsiklauri, D., Viollier, R. D. 1998, Prog. Nucl. Part. Phys.
40, 17
\reference{}
Bili\'c. N.  and Viollier, R. D. 1998, Nucl. Phys. (Proc. Suppl.), B66, 256
\reference{}
Braginsky, V. B., Gurevich, A. V. and Zybin, K. P. 1992, Phys. Lett., A171,
 275
\reference{}
Boehm, F. and Vogel, P. 1987,  Physics of Massive Neutrinos
 (Cambridge: Cambridge Univ. Press ), 76
\reference{}
Eckhardt, D. H. 1993, Phys. Rev., D48, 3762
\reference{}
Freese, K., Fields, B. and Graff, D.,  astro-ph/9901178

\reference{}
Ghez, A. M., Klein, B. L., Morris, M., and Becklin, E. E. 1998, ApJ, 509,
678
\reference{}
Gr\o n,\O . and Soleng, H. H. (1996), ApJ, 456,445

\reference{}
Hammond, R. 1994, in  Matters of Gravity (Electronic Newsletter),
ed. J. Pullin, Vol. 3,  25
\reference{}
Jungman, G., Kamionkowski, M. and Griest, K. 1996, Phys. Rep. 267, 195
\reference{}
Katz, J. I., gr-qc/9809070
\reference{}
Khloper, M. Y. et al. 1991, AZh, 68, 45
\reference{}
Landau, L. and Lifshitz, D. 1960,  Mechanics (Oxford: Pergamon Press)
\reference{}
Macchetto, F. {\it et al.} 1997, ApJ, 489,579 
\reference{}
Munyaneza, F., Tsiklauri, D.,  and Viollier, R. D. 1998, ApJL, 509, L105
\reference{}
Munyaneza, F., Tsiklauri, D.,  and Viollier, R. D. 1999, ApJ, 526, to be published,
astro-ph/9903342
\reference{}
Munyaneza, F. and Viollier, R. D., MNRAS, submitted, astro-ph/9907318
\reference{}
Murphy, E. M. , gr-qc/9810015
\reference{}
Milgrom, M. 1983, ApJ, 270, 365
\reference{}
Mikkelsen, D. R. and Newman, M. J. 1977, Phys. Rev., D16, 919
\reference{}
Nieto, M. M. and Goldman, T. 1992 ,Phys. Rep., 216, 343
\reference{}
Oort, J. H. 1965,  in  Stars and Stellar Systems, Vol 5,
Galactic Structure, ed. A. Blaauw and M. Schmidt (Chicago: University of
Chicago Press),  455
\reference{}
Sanders, R. H. 1986, A\&A, 154, 135 
\reference{}
Sanders, R. H. 1990, Astron. Astrophys. Rev., 2, 1
\reference{}
Sato, O. 1999, CHORUS results, Nucl. Phys. B ( Proc. Suppl.), 77, 220
\reference{}
Trimble, V. 1987, ARA\&A, 25, 425
\reference{}
Tremaine, S.  1990, in  Baryonic Dark Matter, ed. D.  Lynden-Bell and
G. Gilmore, ( Boston: Klewer Academic Publishers), 37
\reference{}
Tremaine, S. 1992, Physics Today, 45,28
\reference{}
Tsiklauri, D. and Viollier, R. D. 1998, ApJ, 500, 591
\reference{}
Tsiklauri, D. and Viollier, R. D. 1999, Astropart. Phys., in press,
 astro-ph/9805272
\reference{}
Turyshev, S. G., Anderson, J. D. , Laing, A. D., Lau, E., Liu, A. S., and
Nieto, M. N.  1999, gr-qc/9903024 
\reference{}
Viollier, R. D., Leimgruber, F. R., and Trautmann, D. 1992, Phys. Lett.,
B297, 132
\reference{}
Viollier, R. D., Trautmann, D., and Tupper, G. B. 1993, Phys. Lett., B306,
79
\reference{}
Viollier, R. D. 1994, Prog. Part. Nucl. Phys., 32,51
\reference{}
Weinberg, S.  1972, Gravitation and Cosmology
(New York: John Wiley and Sons) 
\end{references}
\end{document}